\documentclass[runningheads]{llncs}

\usepackage[T1]{fontenc}
\usepackage{graphicx}
\usepackage{url}

\begin{document}

\title{From Tools to Teacher-Built Teammates: No-Code Pedagogical Plugin Authoring with LearnAdapt Agentic Studio and PedOS 1.1 Lumina}

\titlerunning{Teacher-Built Teammates with Agentic Studio and PedOS}

\author{Nizam Kadir}

\authorrunning{N. Kadir}

\institute{Singapore University of Technology and Design\\
\email{nizam\_kadir@mymail.sutd.edu.sg}}

\maketitle

\begin{abstract}
Teachers and researchers need to adapt educational AI to local goals, but most systems remain difficult to customize or study without coding expertise. We present LearnAdapt Agentic Studio on PedOS 1.1 Lumina, a no-code authoring and governed runtime environment for educational AI plugins. A non-coder describes a desired learning interaction in plain English; the system prepares a previewable plugin artifact, runs safety checks, and supports submission for review. PedOS then deploys approved plugins into a directory for installation. Crucially, telemetry is strictly gated to authenticated users running approved plugins. The demo shows the complete lifecycle from prompt to governed evidence capture, shifting from fixed tools to teacher-built teammates.

\end{abstract}

\section*{Demo Link}

\url{https://www.youtube.com/watch?v=PjIRnAKUxWc}

\section{Motivation}

The AIED 2026 theme raises a practical question: who shapes the AI teammate? \cite{holmes2019}. In many deployments, educators are unable to author learning interactions \cite{molenaar2022} or control evidence capture \cite{unesco2023}, remaining users of fixed systems \cite{zawacki2019}. LearnAdapt addresses this by treating AI functionality as governable plugins. A non-coder describes an activity in plain English, previews the plugin, and submits it for review. After approval, it is distributed via PedOS 1.1 Lumina. This design ensures pedagogical utility, enforces safety reviews, and captures research evidence through explicit, controlled telemetry pathways.

\section{System Presentation}

LearnAdapt Agentic Studio and PedOS 1.1 Lumina form a no-code authoring and governed runtime pipeline.

\textbf{Agentic Studio} is the creator interface where an educator's prompt acts as a client brief for an agent development team. The workflow begins with deterministic preflight checks, followed by specialized agents handling pedagogy, UX, and architecture in parallel. Next, build and evidence design occur simultaneously, culminating in QA, preview iteration, and release packaging. This multi-agent pipeline generates runtime PHP/JS/CSS artifacts that execute securely via platform routing. Direct access is blocked, and creators preview interactions before admin review.

\begin{figure}[htbp]
    \centering
    \includegraphics[width=0.95\linewidth]{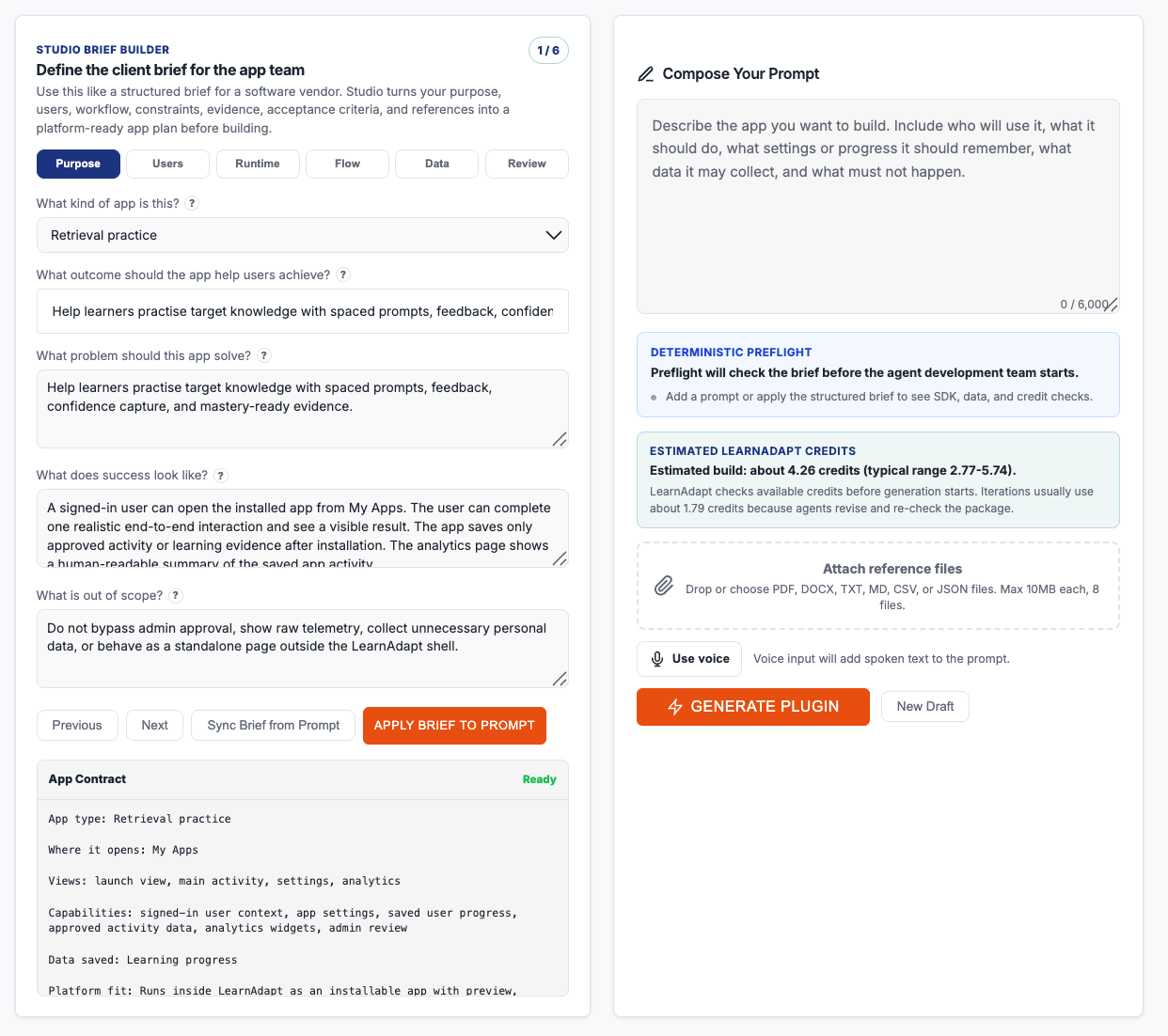}
    \caption{The LearnAdapt Agentic Studio authoring interface. Educators describe the desired learning interaction in plain English, and the agentic pipeline generates a deployable, governable plugin through transparent build stages.}
    \label{fig:agentic-studio}
\end{figure}

\textbf{Security and review} forms the governance layer. Validation includes PHP syntax, API boundaries, and static application security testing (SAST). Admin approval re-runs these checks before deployment, using SAST scores alongside policy status.

\textbf{PedOS 1.1 Lumina} is the runtime layer. Approved plugins are versioned and surfaced in the Plugin Directory for users to preview, install, and access. Strict runtime safeguards isolate plugin execution to prevent platform-wide failures.

\textbf{Telemetry gating} links interaction to evidence. Preview modes use mock telemetry. Events are persisted only for authenticated users running active, installed plugins via approved endpoints.

\section{Demonstrated Functionality}

The demo presents a full plugin lifecycle. An educator prompts Agentic Studio for a science retrieval-practice plugin. After previewing the generated artifact, they submit it. An admin reviewer inspects the artifact, preview, SAST score, and policy status, then approves it. The Plugin Directory displays the deployed package, allowing learners to install and use it. The installed plugin emits telemetry only through approved endpoints. This demonstrates how educators can create bounded, inspectable AI interactions (e.g., misconception probes). Teachers define pedagogical intent; Agentic Studio handles technical production; PedOS governs distribution and evidence capture.

\section{Originality, Maturity, and Future Work}

The main contribution combines no-code AI authoring with enforceable educational governance. LearnAdapt unites prompt-based creation, SAST validation, admin review, deployment, and gated telemetry into one workflow. The system is validated across production paths, demonstrating prompt intake to secure telemetry persistence. While this submission does not claim learning gains, its contribution is a working system for teacher-authored AI. Next steps include a June 2026 co-design workshop to study teachers' prioritized plugin categories and evidence requirements.

\end{document}